# Self-Dual Non-Abelian $N=1$ Tensor Multiplet in $D=2+2$ Dimensions

Hitoshi Nishino[1]) and Subhash Rajpoot[2])

*Department of Physics & Astronomy*
*California State University*
*1250 Bellflower Boulevard*
*Long Beach, CA 90840*

## Abstract

We present a self-dual non-Abelian $N = 1$ supersymmetric tensor multiplet in $D = 2 + 2$ space-time dimensions. Our system has three on-shell multiplets: (i) The usual non-Abelian Yang-Mills multiplet $(A_\mu{}^I, \lambda^I)$ (ii) A non-Abelian tensor multiplet $(B_{\mu\nu}{}^I, \chi^I, \varphi^I)$, and (iii) An extra compensator vector multiplet $(C_\mu{}^I, \rho^I)$. Here the index $I$ is for the adjoint representation of a non-Abelian gauge group. The duality symmetry relations are $G_{\mu\nu\rho}{}^I = -\epsilon_{\mu\nu\rho}{}^\sigma \nabla_\sigma \varphi^I$, $F_{\mu\nu}{}^I = +(1/2)\,\epsilon_{\mu\nu}{}^{\rho\sigma} F_{\rho\sigma}{}^I$, and $H_{\mu\nu}{}^I = +(1/2)\,\epsilon_{\mu\nu}{}^{\rho\sigma} H_{\rho\sigma}{}^I$, where $G$ and $H$ are respectively the field strengths of $B$ and $C$. The usual problem with the coupling of the non-Abelian tensor is avoided by non-trivial Chern-Simons terms in the field strengths $G_{\mu\nu\rho}{}^I$ and $H_{\mu\nu}{}^I$. For an independent confirmation, we re-formulate the component results in superspace. As applications of embedding integrable systems, we show how the $\mathcal{N} = 2$, $r = 3$ and $\mathcal{N} = 3$, $r = 4$ flows of generalized Korteweg-de Vries equations are embedded into our system.

[1]) E-Mail: hnishino@csulb.edu
[2]) E-Mail: rajpoot@csulb.edu

## 1. Introduction

Considerable progress has been achieved in constructing theories with consistent interactions of non-Abelian tensor fields of 2nd-rank or higher [1][2][3][4]. The key ingredient is based on the so-called 'vector-tensor hierarchies' [1][2][3][4], utilizing extra Chern-Simons (CS) terms added to the field strengths of non-Abelian tensors. Another important technique is the engagement of generalized Stueckelberg formalism for higher-rank tensors, avoiding the usual inconsistency of non-Abelian tensor couplings.

In [1], the gauging of five-dimensional (5D) maximal supergravity with $E_{6(+6)}$ was generalized in terms of the so-called 'vector-tensor hierarchy'. The field strength $\mathcal{H}_{\mu\nu\rho\,I}$ for a 2nd-rank antisymmetric tensor $B_{\mu\nu\,I}$ is introduced with generalized CS terms [1], such that $\mathcal{H}_{\mu\nu\rho\,I}$ is invariant under tensor and vector gauge transformations. Subsequently, the relationship of the formulation in [1] with M-theory [5][6] was confirmed by representation assignments [2]. Applications to gauged maximal supergravity in 3D were also performed with all possible tensor fields [2]. It is suggested in [3] that the system of the non-Abelian gauge group $G \times G$ fits nicely to multiple M5-branes with manifest (1,0) supersymmetry. In [4], the original vector-tensor hierarchy was simplified further to 'minimal vector-tensor hierarchy' in the context of conformal $N = (1,0)$ supergravity in 6D.

Motivated by this series of developments [1][2][3][4], we have presented in our previous paper [7] an $N = 1$ supersymmetric formulation of non-Abelian tensor in 4D. Our formulation is understood as a special case of the so-called minimal vector-tensor hierarchy [4]. Our field strengths are the tensor multiplet (TM) $(B_{\mu\nu}{}^I, \chi^I, \varphi^I),^{3)}$ the Yang-Mills vector multiplet (YMVM) $(A_\mu{}^I, \lambda^I)$ and the extra compensating vector multiplet (ECVM) $(C_\mu{}^I, \rho_\mu{}^I)$. Following the 'vector-tensor hierarchy' [1][2][4], we define our field strengths by [7]

$$F_{\mu\nu}{}^I \equiv +2\partial_{[\mu} A_{\nu]}{}^I + g f^{IJK} A_\mu{}^J A_\nu{}^K \ , \tag{1.1a}$$

$$G_{\mu\nu\rho}{}^I \equiv +3D_{[\mu} B_{\nu\rho]}{}^I - 3 f^{IJK} C_{[\mu}{}^J F_{\nu\rho]}{}^K \ , \tag{1.1b}$$

$$H_{\mu\nu}{}^I \equiv +2D_{[\mu} C_{\nu]}{}^I + g B_{\mu\nu}{}^I \ . \tag{1.1c}$$

---

3) We use the indices $\mu, \nu, \cdots = 0, 1, 2, 3$ for the space-time coordinates.

Relevantly, these field strengths satisfy Bianchi identities

$$D_{[\mu} F_{\nu\rho]}{}^I \equiv 0 \quad , \tag{1.2a}$$

$$D_{[\mu} G_{\nu\rho\sigma]}{}^I \equiv +\tfrac{3}{2} f^{IJK} F_{[\mu\nu}{}^J H_{\rho\sigma]}{}^K \quad , \tag{1.2b}$$

$$D_{[\mu} H_{\nu\rho]}{}^I \equiv +\tfrac{1}{3} g \, G_{\mu\nu\rho}{}^I \quad . \tag{1.2c}$$

Due to the indices $\mu\nu$ on $B_{\mu\nu}{}^I$ or $\mu$ on $C_\mu{}^I$, there should be also proper gauge transformations for these fundamental fields. Let us call them $\delta_\beta$ and $\delta_\gamma$-gauge transformations. In addition to the YM gauge transformation $\delta_\alpha$, their explicit forms are

$$\delta_\alpha(A_\mu{}^I, \; B_{\mu\nu}{}^I, \; C_\mu{}^I) = (+D_a\mu^I, \; -f^{IJK}\alpha^J B_{\mu\nu}{}^K, \; -f^{IJK}\alpha^J C_\mu{}^K) \quad , \tag{1.3a}$$

$$\delta_\beta(A_\mu{}^I, \; B_{\mu\nu}{}^I, \; C_\mu{}^I) = (0, \; +2D_{[\mu}\beta_{\nu]}{}^I, \; -g\beta_\mu{}^I) \quad , \tag{1.3b}$$

$$\delta_\gamma(A_\mu{}^I, \; B_{\mu\nu}{}^I, C_\mu{}^I) = (0, \; -f^{IJK}F_{\mu\nu}{}^J\gamma^K, \; D_\mu\gamma^I) \quad . \tag{1.3c}$$

As (1.1c) or (1.3b) shows, $C_\mu{}^I$ is a vectorial Stueckelberg field, absorbed into the longitudinal component of $B_{\mu\nu}{}^I$. Due to the general hierarchy [1][2][4], all field strengths are covariant under $\delta_\alpha$ and invariant under $\delta_\beta$ and $\delta_\gamma$:

$$\delta_\alpha(F_{\mu\nu}{}^I, \; G_{\mu\nu\rho}{}^I, \; H_{\mu\nu}{}^I) = -f^{IJK}\alpha^J(F_{\mu\nu}{}^K, \; G_{\mu\nu\rho}{}^K, \; H_{\mu\nu}{}^K) \quad , \tag{1.4a}$$

$$\delta_\beta(F_{\mu\nu}{}^I, \; G_{\mu\nu\rho}{}^I, \; H_{\mu\nu}{}^I) = 0 \quad , \quad \delta_\gamma(F_{\mu\nu}{}^I, \; G_{\mu\nu\rho}{}^I, \; H_{\mu\nu}{}^I) = 0 \quad . \tag{1.4b}$$

In the present paper, we apply these developments [1][2][3][4][7] to 'self-dual tensor multiplets' in $2+2$ dimensions ($D = 2+2$). The original 'self-duality' was implied in terms of Hodge-Poincaré duality, applied to self-dual Yang-Mills (SDYM) theory [8][9]. There are two grounds for the importance of SDYM theory [8][9]. First, it has been known that $N = 2$ superstring requires the background YM field be self-dual in $D = 2+2$ space-time dimensions [10]. Second, SDYM theory seems to be the 'master theory' of all (bosonic) integrable models in lower dimensions $1 \leq D \leq 3$ [9]. The supersymmetrization of SDYM, *i.e.*, self-dual supersymmetric Yang-Mills (SDSYM) theory was also accomplished in 1990's [11][12][13]. In particular, the maximally supersymmetric SDSYM theory in $D = 2+2$ is $N = 8$ case [14].

From a naïve viewpoint in the context of SDSYM, there appears to be *no* strong motivation to consider *tensor* fields carrying non-Abelian indices. Because there are three major objections against such a trial. First, the original conjecture [9] was about SDYM fields, that may generate all the integrable models in lower dimensions. So an additional tensor field seems redundant. Second, even for $N = 1$ superstring theory [15], a 2-form tensor field background should carry *no* additional indices, so that a non-Abelian tensor seems to be irrelevant. Third, even independent of string theory [15], it is not interesting enough, unless the tensor carries non-trivial indices such as adjoint index with non-trivial interactions. On the other hand, non-Abelian tensor couplings to a YM field used to be problematic, before the non-Abelian tensor formulations, such as [1][2][3][4][7] were established.

Aforementioned three objections, however, are considered obsolete nowadays. Definitely, the first objection seems invalid, since the duality symmetry between the 3-form field strength $G_{\mu\nu\rho}$ and the 1-form field strength $\nabla_\mu\varphi$ of a dilaton was predicted as important backgrounds for $N = (2, 0)$ heterotic $\sigma$-model [10]. The second objection is not strong enough to avoid the discussion of non-Abelian tensor with duality and supersymmetry. Because even if tensors with additional indices may not be directly related to $N = 1$ [15] or $N = 2$ [10] superstring, duality symmetry between a 3-form and 1-form field strengths [12] may well be associated with integrable models in lower dimensions. The third objection has also lost its strong ground, because of the above-mentioned breakthrough [1][2][3][4][7]. Moreover, important relationships between vector-tensor hierarchy and M-theory [5] have been also established in [2].

Motivated by these viewpoints, especially by the success of the supersymmetrization of non-Abelian tensor [7], we give in the present paper the component formulation [16] of self-dual non-Abelian tensor multiplet (SDNATM)[4)] in $D = 2+2$. There are three multiplets in our system: (i) The usual non-Abelian YM vector multiplet (VM) $(A_\mu{}^I, \lambda^I)$, (ii) A SDNATM

---

[4)] The original tensor (or linear) multiplet *without* self-duality was first formulated in [17]. Here we deal with 'self-dual' NATM. The tensor $B_{\mu\nu}{}^I$ itself in this multiplet is *not* self dual. However, since the scalar $\varphi^I$ and $B_{\mu\nu}{}^I$ within NATM are dual to each other, we call this multiplet as a 'self-dual' NATM.

$(B_{\mu\nu}{}^I, \chi_\mu{}^I, \varphi^I)$, and (iii) An ECVM $(C_\mu{}^I, \rho^I)$. Our duality conditions are[5)]

$$G_{\mu\nu\rho}{}^I \overset{*}{=} -\epsilon_{\mu\nu\rho}{}^\sigma \nabla_\sigma \varphi^I \ , \tag{1.5a}$$

$$\nabla_\mu \varphi^I \overset{*}{=} +\tfrac{1}{6}\epsilon_\mu{}^{\nu\rho\sigma} G_{\rho\sigma\tau}{}^I \ , \tag{1.5b}$$

$$H_{\mu\nu}{}^I \overset{*}{=} +\tfrac{1}{2}\epsilon_{\mu\nu}{}^{\rho\sigma} H_{\rho\sigma}{}^I \ , \tag{1.5c}$$

$$F_{\mu\nu}{}^I \overset{*}{=} +\tfrac{1}{2}\epsilon_{\mu\nu}{}^{\rho\sigma} F_{\rho\sigma}{}^I \ . \tag{1.5d}$$

Eqs. (1.5a) and (1.5b) imply the Hodge-Poincaré duality symmetry between the two field strengths $G_{\mu\nu\rho}{}^I$ and $\nabla_\mu\varphi^I$, while (1.5c) and (1.5d) are the usual SD for the field strengths $H$ and $F$. The Abelian case without the adjoint index has been well known for a while [12]. However, the new ingredient here is that the self-dual TM carrying the adjoint index of a non-Abelian gauge group, and we have to accomplish the consistent couplings between the tensor field and the usual YM gauge field, following the vector-tensor hierarchies [1][2][4][7].

As a general feature of SDYM systems. it has been well known that SDYM theory lacks an action, unless one breaks Lorentz invariance [18]. This can be easily understood as follows. If we try to construct the kinetic term of a self-dual field strength $F_{\mu\nu}{}^I$, it will be a total divergence:

$$-\tfrac{1}{4} F_{\mu\nu}{}^I F^{\mu\nu\, I} \overset{*}{=} -\tfrac{1}{4}\left(\tfrac{1}{2}\epsilon_{\mu\nu}{}^{\rho\sigma} F_{\rho\sigma}{}^I\right) F^{\mu\nu\, I} = -\tfrac{1}{8}\epsilon^{\mu\nu\rho\sigma} F_{\mu\nu}{}^I F_{\rho\sigma}{}^I \overset{\triangledown}{=} 0 \ , \tag{1.6}$$

where $\overset{\triangledown}{=}$ is an equality up to a total divergence. This is also confirmed by varying $A_\mu{}^I$ in (1.6) with a zero result, due to the Bianchi identity $D_{[\mu} F_{\nu\rho]}{}^I \equiv 0$. Another typical example is self-dual 5-th rank field strength in the so-called type IIB supergravity in 10D [19]. This property is shared also with the duality symmetric field strengths $G_{\mu\nu\rho}{}^I$ and $D_\mu\varphi^I$ satisfying (1.5a) and (1.5b):

$$
\begin{aligned}
-\tfrac{1}{12} G_{\mu\nu\rho}{}^I G^{\mu\nu\rho\, I} &\overset{*}{=} -\tfrac{1}{12}\left(-\epsilon_{\mu\nu\rho}{}^\sigma D_\sigma\varphi^I\right) G^{\mu\nu\rho\, I} \\
&= +\tfrac{1}{12}\epsilon^{\mu\nu\rho\sigma} G_{\mu\nu\rho}{}^I D_\sigma\varphi^I \overset{\triangledown}{=} +\tfrac{1}{12}\epsilon^{\mu\nu\rho\sigma}\varphi^I D_{[\mu} G_{\nu\rho\sigma]}{}^I \\
&\equiv +\tfrac{1}{8}\epsilon^{\mu\nu\rho\sigma} f^{IJK}\varphi^I F_{\mu\nu}{}^J H_{\rho\sigma}{}^K \ ,
\end{aligned}
\tag{1.7}
$$

---

[5)] We use the symbol $\overset{*}{=}$ for an equality associated with dualities, or ansätze for DRs in section 5. The derivative $\nabla_\mu \equiv \partial_\mu + g A_\mu{}^I T^I$ is YM non-Abelian group covariant. The definitions of these field strengths are the same as (1.1). The notation for the $D = 2+2$ space-time is the same as in [12], such as $\gamma_{\mu\nu\rho\sigma} = +\epsilon_{\mu\nu\rho\sigma}\gamma_5$, $\epsilon_{\mu\nu}{}^{\rho\sigma}\gamma_{\rho\sigma} = -2\gamma_5\gamma_{\mu\nu}$.

where use is made of the Bianchi identity (1.5b). Even though the last side of (1.7) is *not* vanishing, since it is already at the *trilinear* interaction, this can *no* longer regarded as the *kinetic* term.

In order to overcome this general problem with SD field strengths, there have been some methods developed, such as using harmonic superspace [20]. However, we do not attempt to solve the action problem in this paper, regarding it as a separate issue. So instead of giving an explicit lagrangian, we use only the set of field equations.

As applications of SDNATM, we also show some examples of our system generating $\mathcal{N} = 2$, $r = 3$ and $\mathcal{N} = 3$, $r = 4$ flows of generalized Korteweg-de Vries (KdV) equations in $D = 1+1$ [21]. Compared with the case of SDYM system [22][23], our system is relatively simpler, but it still maintains non-trivial feature of embeddings. This seems to be the role played by the TM, showing the advantage of our SDNATM system over the SDSYM system [24].

This paper is organized as follows. In the next section, we will give first the component formulation for SDNATM. In section 3, we will give the superspace re-formulation of the component results. In section 4, we mention the difficulty with *off-shell* formulation in terms of prepotentials and auxiliary fields. In section 5, we will give the embedding of KdV equations in $D = 1 + 1$, as an important application of SDNATM. In section 6, we point it out that bosonic conditions arising in our system after a dimensional reduction (DR) into $D = 1 + 1$ are equivalent to bosonic equations arising in $N = 2$ supersymmetric SDSYM theory [12]. The concluding remarks will be given in section 7, with potential generalizations to higher space-time dimensions.

## 2. Component Formulation

We first give our results in component language in the most conventional notation. In the next section, we will perform the re-formulation in superspace in order to confirm the total consistency.

The dualities and their supersymmetric partner conditions of our system are summarized as (1.5) and the chiralities of fermionic fields

$$\gamma_5 \left( \lambda^I, \, \chi^I, \, \rho^I \right) = \left( -\lambda^I, \, +\chi^I, \, -\rho^I \right) \quad , \tag{2.1}$$

As in the SDSYM case, these chiralities and SD are closely related to each other for the SDNATM system.

The global $N = 1$ supersymmetry transformation rule is

$$\delta_Q A_\mu{}^I = + (\overline{\varepsilon}\gamma_\mu \lambda^I) \quad , \tag{2.2a}$$

$$\delta_Q \lambda^I = + \tfrac{1}{4} (\gamma^{\mu\nu}\epsilon) F_{\mu\nu}{}^I \quad , \tag{2.2b}$$

$$\delta_Q B_{\mu\nu}{}^I = + (\overline{\varepsilon}\gamma_{\mu\nu}\chi^I) \quad , \tag{2.2c}$$

$$\delta_Q \chi^I = + \tfrac{1}{12} (\gamma^{\mu\nu\rho}\epsilon) G_{\mu\nu\rho}{}^I - \tfrac{1}{2}(\gamma^c \epsilon) D_\mu \varphi^I + f^{IJK}\epsilon (\overline{\lambda}^J \rho^K) \quad , \tag{2.2d}$$

$$\delta_Q \varphi^I = + (\overline{\varepsilon}\chi^I) \quad , \tag{2.2e}$$

$$\delta_Q C_\mu{}^I = + (\overline{\varepsilon}\gamma_\mu \rho^I) \quad , \tag{2.2f}$$

$$\delta_Q \rho^I = + \tfrac{1}{4} (\gamma^{\mu\nu}\epsilon) H_{\mu\nu}{}^I \quad . \tag{2.2g}$$

The first and second terms in the r.h.s. of (2.2d) are the same under the duality (1.5a). The consistency between these rules and the dualities (1.5) or chiralities (2.1) will be confirmed later.

The field equations for the fermionic fields are

$$\displaystyle{\not{D}}\lambda^I \doteq 0 \quad , \tag{2.3a}$$

$$\displaystyle{\not{D}}\rho^I - 2g\chi^I \doteq 0 \quad , \tag{2.3b}$$

$$\displaystyle{\not{D}}\chi^I - \tfrac{1}{4} f^{IJK} (\gamma^{\mu\nu}\lambda^J) H_{\mu\nu}{}^K + \tfrac{1}{4} f^{IJK} (\gamma^{\mu\nu}\rho^J) F_{\mu\nu}{}^K + g f^{IJK}\lambda^J \varphi^K \doteq 0 \quad . \tag{2.3c}$$

Additional useful transformation rules are

$$\delta_Q F_{\mu\nu}{}^I = - 2(\overline{\varepsilon}\gamma_{[\mu}D_{\nu]}\lambda^I) \quad , \tag{2.4a}$$

$$\delta_Q G_{\mu\nu\rho}{}^I = + 3(\overline{\varepsilon}\gamma_{[\mu\nu}D_{\rho]}\chi^I) + 3f^{IJK}(\overline{\varepsilon}\gamma_{[\mu}\lambda^J) H_{\nu\rho]}{}^K + 3f^{IJK}(\overline{\varepsilon}\gamma_{[\mu}\rho^J) F_{\nu\rho]}{}^K \quad , \tag{2.4b}$$

$$\delta_Q H_{\mu\nu}{}^I = - 2(\overline{\varepsilon}\gamma_{[\mu}D_{\nu]}\rho^I) + (\overline{\varepsilon}\gamma_{\mu\nu}\chi^I) \quad , \tag{2.4c}$$

The fermionic field equations (2.3) with the chiralities (2.1) are consistent with these transformation rules. For example, we can confirm that

$$0 \overset{?}{=} \delta_Q \left( F_{\mu\nu}{}^I - \tfrac{1}{2} \epsilon_{\mu\nu}{}^{\rho\sigma} F_{\rho\sigma}{}^I \right) = -(\overline{\epsilon}\gamma_{\mu\nu} \slashed{D}\lambda^I) \doteq 0 \quad (Q.E.D.) \quad , \tag{2.5a}$$

$$0 \overset{?}{=} \delta_Q \left( G_{\mu\nu\rho}{}^I + \epsilon_{\mu\nu\rho}{}^\sigma D_\sigma \varphi^I \right) = + \Big[ \overline{\epsilon}\gamma_{\mu\nu\rho} \Big\{ \slashed{D}\chi^I - \tfrac{1}{4} f^{IJK} (\gamma^{\sigma\tau}\lambda^J) H_{\sigma\tau}{}^K$$
$$+ \tfrac{1}{4} f^{IJK} (\gamma^{\sigma\tau}\rho^J) F_{\sigma\tau}{}^K + g f^{IJK} \lambda^J \varphi^K \Big\} \Big] \doteq 0 \quad (Q.E.D.) \quad , \tag{2.5b}$$

$$0 \overset{?}{=} \delta_Q \left( H_{\mu\nu}{}^I - \tfrac{1}{2} \epsilon_{\mu\nu}{}^{\rho\sigma} H_{\rho\sigma}{}^I \right) = - \Big[ \overline{\epsilon}\gamma_{\mu\nu} \left( \slashed{D}\rho^I - 2g\chi^I \right) \Big] \doteq 0 \quad (Q.E.D.) \quad . \tag{2.5c}$$

In these confirmations, use is made of the $\gamma$-matrix algebra, such as

$$\epsilon_{\mu\nu\rho}{}^\sigma \gamma_\sigma = \gamma_5 \gamma_{\mu\nu\rho} = -\gamma_{\mu\nu\rho}\gamma_5 \quad , \quad \{\gamma_{\mu\nu\rho}, \, \gamma^{\sigma\tau}\} = +12\delta_{[\mu|}{}^\sigma \delta_{|\nu|}{}^\tau \gamma_{|\rho]} \quad . \tag{2.6}$$

## 3. Superspace Re-Formulation

We have so far presented only component formalism. Even though we have performed cross-confirmations such as (2.5), it is still better to have independent confirmation in superspace. To this end, we use the superspace notations, such as the indices $A = (a, \alpha, \dot{\alpha}), B = (b, \beta, \dot{\beta}), \cdots$ for the superspace coordinates, where $a, b, \cdots = 0, 1, 2, 3$ (or $\alpha, \beta, \cdots = 1, 2, 3, 4; \dot{\alpha}, \dot{\beta}, \cdots = \dot{1}, \dot{2}, \dot{3}, \dot{4}$) are for the bosonic (or fermionic) coordinates. As usual the *undotted* (or *dotted*) indices are for the chiral (or ant-chiral) fermions. Accordingly, our field content in superspace notation is VM $(A_a{}^I, \overline{\lambda}_{\dot{\alpha}}{}^I)$, SDNATM $(B_{ab}{}^I, \chi_\alpha{}^I, \varphi^I)$ and ECVM $(C_a{}^I, \overline{\rho}_{\dot{\alpha}}{}^I)$. Our (anti)symmetrizations are such as $X_{[AB)} \equiv X_{AB} - (-1)^{AB} X_{BA}$ *without* the factor of $1/2$.

The *off-shell* superspace formulation of the original linear multiplet [17] has been systematically studied [25]. In off-shell formulations, the so-called prepotentials drastically simplify the total system. Even though we know that such off-shell formulation is much more advantageous than *on-shell* formulation, we do not have a complete off-shell formulation for *non-Abelian* tensor multiplets at the present time, even in the usual $D = 3 + 1$ space-time [7]. For this reason, we do not attempt to give the off-shell formulation in superspace in this paper. Instead we use the Bianchi identities in superspace, as a guiding principle for our *on-shell* formulation.

Based on this principle, following the definitions of the $F$, $G$ and $H$-field strengths in the component formulation (1.1), our corresponding superspace definitions are

$$F_{AB}{}^I \equiv + \nabla_{[A} A_{B)}{}^I - T_{AB}{}^C A_C{}^I + g f^{IJK} A_A{}^J A_B{}^K \quad , \tag{3.1a}$$

$$G_{ABC}{}^I \equiv + \tfrac{1}{2} \nabla_{[A} B_{CD)}{}^I - \tfrac{1}{2} T_{[AB|}{}^D B_{D|C)}{}^I - \tfrac{1}{2} f^{IJK} C_{[A}{}^J F_{BC)}{}^K \quad , \tag{3.1b}$$

$$H_{AB}{}^I \equiv + \nabla_{[A} C_{B)}{}^I - T_{AB}{}^C B_C{}^I + g B_{AB}{}^I \quad . \tag{3.1c}$$

Correspondingly, our superspace Bianchi identities (BIds) for these superfield strengths are

$$+ \tfrac{1}{2} \nabla_{[A} F_{BC)}{}^I - \tfrac{1}{2} T_{[AB|}{}^D F_{D|C)}{}^I \equiv 0 \quad , \tag{3.2a}$$

$$+ \tfrac{1}{6} \nabla_{[A} G_{BCD)}{}^I - \tfrac{1}{4} T_{[AB|}{}^E G_{E|CD)}{}^I - \tfrac{1}{4} f^{IJK} F_{[AB}{}^J H_{CD)}{}^K \equiv 0 \quad , \tag{3.2b}$$

$$+ \tfrac{1}{2} \nabla_{[A} H_{BC)}{}^I - \tfrac{1}{2} T_{[AB|}{}^D H_{D|C)}{}^I - g G_{ABC}{}^I \equiv 0 \quad . \tag{3.2c}$$

These are nothing but the superspace generalization of the component case (1.2). These are also parallel to the non-self-dual formulation in $D = 3 + 1$ [7]. Since we have the corresponding non-dual case in $D = 3 + 1$, even though our formulation is *on-shell* formulation without prepotentials, the comparison with the $D = 3 + 1$ case [7] is straightforward.

There are, however, differences in $D = 2 + 2$ about chiralities of spinors, compared with $D = 3 + 1$ in [7]. A special treatment is needed for spinors in $D = 2 + 2$ [11][12][13]. The most important feature is that *dotted* spinors are *independent* of *un-dotted* spinors. This situation is different from the case of $D = 3 + 1$ [7], where *dotted* spinors are just complex conjugate to *un-dotted* spinors [26]. This gives certain differences compared with our result in $D = 3 + 1$ [7].

Our superspace constraints at the engineering dimensions $0 \leq d \leq 1$ are

$$T_{\alpha\dot\beta}{}^c = + (\gamma^c)_{\alpha\dot\beta} \quad , \qquad G_{\alpha\dot\beta c}{}^I = + (\gamma_c)_{\alpha\dot\beta} \varphi^I \quad , \tag{3.3a}$$

$$G_{\alpha bc}{}^I = - (\gamma_{bc} \chi^I)_\alpha \quad , \qquad F_{\alpha b}{}^I = - (\gamma_b \overline\lambda{}^I)_\alpha \quad , \qquad H_{\alpha b}{}^I = - (\gamma_b \overline\rho{}^I)_\alpha \quad , \tag{3.3b}$$

$$\nabla_\alpha \varphi = - \chi_\alpha{}^I \quad , \tag{3.3c}$$

$$\overline\nabla_{\dot\alpha} \overline\lambda_{\dot\beta}{}^I = + \tfrac{1}{4} (\gamma^{cd})_{\alpha\beta} F_{ab}{}^I \quad , \tag{3.3d}$$

$$\overline{\nabla}_{\dot\alpha}\chi_\beta{}^I = -\tfrac{1}{12}(\gamma^{cde})_{\dot\alpha\beta}G_{cde}{}^I - \tfrac{1}{2}(\gamma^c)_{\beta\dot\alpha}\nabla_c\varphi^I \overset{*}{=} -(\gamma^c)_{\beta\dot\alpha}\nabla_c\varphi^I \quad, \tag{3.3e}$$

$$\nabla_\alpha\chi_\beta{}^I = -C_{\alpha\beta}f^{IJK}(\overline{\lambda}^J\overline{\rho}^K) \quad, \tag{3.3f}$$

$$\overline{\nabla}_{\dot\alpha}\overline{\rho}_{\dot\beta}{}^I = +\tfrac{1}{4}(\gamma^{cd})_{\dot\alpha\dot\beta}H_{cd}{}^I + gC_{\dot\alpha\dot\beta}\varphi^I \quad. \tag{3.3g}$$

In (3.3e), the last equality is valid under the duality symmetry (1.5a). All other constraints with independent components, such as $\nabla_\alpha\overline{\rho}_{\dot\beta}{}^I$ or $\overline{\nabla}_{\dot\alpha}\varphi^I$, $etc.$ are all zero. In particular, $\overline{\nabla}_{\dot\alpha}\varphi^I = 0$ implies that $\varphi^I$ is a $chiral$ scalar superfield [12]. These structures are very similar to the TM case in $D = 3 + 1$ in [7]. The only exceptions are such as the absence of fermionic bilinear terms, and coefficients such as those in (2.3a), (2.3d) or (2.3e) are half of the corresponding ones in [7]. These facts are the reflections of the chiral nature of our present system. As usual in superspace, the constraints in (3.3) satisfy the BIds at $0 \leq d \leq 1$.

At dimension $d = 3/2$, BIds (3.2) lead to

$$\nabla_\alpha G_{bcd}{}^I = -\tfrac{1}{2}(\gamma_{[bc}\nabla_{d]}\chi^I)_\alpha - \tfrac{1}{2}f^{IJK}(\gamma_{[b|}\overline{\lambda}^J)_\alpha H_{|cd]}{}^K + \tfrac{1}{2}f^{IJK}(\gamma_{[b|}\overline{\rho}^J)_\alpha F_{|cd]}{}^K \quad, \tag{3.4a}$$

$$\nabla_\alpha H_{bc}{}^I = +(\gamma_{[b}\nabla_{c]}\overline{\rho}^I)_\alpha - g(\gamma_{bc}\chi^I)_\alpha \quad, \qquad \nabla_\alpha F_{bc}{}^I = +(\gamma_{[b}\nabla_{c]}\overline{\lambda}^I)_\alpha \quad, \tag{3.4b}$$

and the fermionic field equations[6)]

$$(\overline{\nabla}\!\!\!/\,\overline{\lambda}^I)_\alpha \overset{.}{=} 0 \quad, \tag{3.5a}$$

$$(\overline{\nabla}\!\!\!/\,\overline{\rho}^I)_\alpha - 2g\chi_\alpha{}^I \overset{.}{=} 0 \quad, \tag{3.5b}$$

$$(\nabla\!\!\!/\,\chi^I)_{\dot\alpha} - \tfrac{1}{4}f^{IJK}(\gamma^{ab}\overline{\lambda}^J)_{\dot\alpha}H_{ab}{}^K + \tfrac{1}{4}f^{IJK}(\gamma^{ab}\overline{\rho}^J)_{\dot\alpha}F_{ab}{}^K + gf^{IJK}\overline{\lambda}_{\dot\alpha}{}^J\varphi^K \overset{.}{=} 0 \quad. \tag{3.5c}$$

Needless to say, these are consistent with the component results (2.3) and (2.4).

Compared with the duality-less case in $D = 3 + 1$ [7], the structures in (3.5) have differences as well as similarities. The similarity is the parallel structure of the constraints (3.3) to [7]. The difference is that our fermionic field equations in (3.5) are much simpler, because of $chirality$ associated with dualities, simplifying or deleting certain terms in these field equations. Compared with the $D = 3 + 1$ case [7], our present system has $no$ higher-order terms that are skipped in [7]. For example, (fermion)$^2$-terms are absent in (3.3d)

---

[6)] We use the symbol $\overset{.}{=}$ for a field equation, or an ansatz for a solution.

and (3.3g), while in $D = 3 + 1$ [7] corresponding terms are present. This is nothing bizarre, considering the fact that each fermion has a definite chirality, so that possible terms are limited. This can be rigorously confirmed in superspace than in component language, because Fierz rearrangements are more transparent.

We next study various self-consistencies of our system. First, we can show the consistency of the anticommutators on the fermions:

$$\{\nabla_\alpha, \overline{\nabla}_{\dot\beta}\} \overline{\rho}_{\dot\gamma}{}^I = + T_{\alpha\dot\beta}{}^c \nabla_c \overline{\rho}_{\dot\gamma}{}^I + \tfrac{1}{2} C_{\dot\beta\dot\gamma} (\overline{\nabla}\overline{\rho}^I - 2g\chi^I)_\alpha \doteq + T_{\alpha\dot\beta}{}^c \nabla_c \overline{\rho}_{\dot\gamma}{}^I \quad, \tag{3.6a}$$

$$\{\nabla_\alpha, \overline{\nabla}_{\dot\beta}\} \chi_\gamma{}^I = + T_{\alpha\dot\beta}{}^c \nabla_c \chi_\gamma{}^I + C_{\alpha\gamma}\Big[ \overline{\nabla}\chi^I - \tfrac{1}{4} f^{IJK}(\gamma^{ab}\overline{\lambda}^J)H_{ab}{}^K$$
$$+ \tfrac{1}{4} f^{IJK}(\gamma^{ab}\overline{\rho}^J)F_{ab}{}^K + g f^{IJK}\overline{\lambda}^J \varphi^K \Big]_{\dot\beta} \doteq + T_{\alpha\dot\beta}{}^c \nabla_c \chi_\gamma{}^I \quad, \tag{3.6b}$$

$$\{\nabla_\alpha, \overline{\nabla}_{\dot\beta}\} \overline{\lambda}_{\dot\gamma}{}^I = + T_{\alpha\dot\beta}{}^c \nabla_c \overline{\lambda}_{\dot\gamma}{}^I + \tfrac{1}{2} C_{\dot\beta\dot\gamma}(\overline{\nabla}\overline{\lambda}^I)_\alpha \doteq + T_{\alpha\dot\beta}{}^c \nabla_c \overline{\lambda}_{\dot\gamma}{}^I \quad, \tag{3.6c}$$

where use is made of the fermionic field equations (3.5).

Second, we can also re-obtain SD (1.5) from the fermionic field equations (3.5). For example, we can re-obtain the SD of $F$ and the $G$-$\nabla\varphi$ duality from the $\overline{\rho}$-field equation:

$$0 \overset{?}{=} (\gamma^a)^{\alpha\dot\beta} \overline{\nabla}_{\dot\beta} \Big[ (\overline{\nabla}\overline{\rho}^I) - 2g\chi^I \Big]_\alpha$$
$$\overset{*}{=} -2g\left(\nabla^a\varphi^I - \tfrac{1}{6}\epsilon^{abcd}G_{bcd}{}^I\right) + \nabla_b\left(H^{abI} - \tfrac{1}{2}\epsilon^{abcd}H_{cd}{}^I\right) \overset{*}{=} 0 \qquad (Q.E.D.) \tag{3.7}$$

The symbol $\overset{*}{=}$ implies that we used the last expression of (3.3e). Eq. (3.7) holds under (1.5b) and (1.5c).

Another example is for the self-dualities of $F$ and $H$-field strengths re-obtained from the $\chi$-field equation:

$$0 \overset{?}{=} (\gamma^{ab})^{\alpha\dot\beta} \overline{\nabla}_{\dot\alpha}\Big[ + (\overline{\nabla}\chi^I) - \tfrac{1}{4}f^{IJK}(\gamma^{cd}\overline{\lambda}^J)H_{cd}{}^K + \tfrac{1}{4}f^{IJK}(\gamma^{cd}\overline{\rho}^J)F_{cd}{}^K + g f^{IJK}\overline{\lambda}^J\varphi^K \Big]_{\dot\beta}$$
$$\overset{*}{=} \tfrac{1}{2}f^{IJK}\left(F_{c[a|}{}^J - \tfrac{1}{2}\epsilon_{c[a|}{}^{de}F_{de}{}^J\right)H_{c|b]}{}^K + \tfrac{1}{2}f^{IJK}F_{c[a|}{}^J\left(H_{c|b]}{}^K - \tfrac{1}{2}\epsilon_{c|b]}{}^{de}H_{de}{}^K\right) \tag{3.8a}$$

$$\overset{*}{=} 0 \qquad (Q.E.D.) \tag{3.8b}$$

The symbol $\overset{*}{=}$ in (3.8a) implies that the last expression of (3.3e) is used. Eq. (3.8b) holds under the SD on $F$ and $H$, as desired.

## 4. Difficulty with Off-Shell Prepotential Formulation

One may wonder, whether we can use 'off-shell' formulation in terms of prepotential superfield. The advantage of off-shell prepotential superfields is that we can compare our results with the conventional system with tensor (linear) multiplets [17][25]. At least in 4D, all the prepotentials for our three multiplets for *Abelian* case have been already known [17][25][26][12].

However, there seems to exist some obstruction against such an idea for *non-Abelian* case. The main problem is caused by the following three features. First, the tensor field carries the non-Abelian adjoint index whose superspace formulation has never been presented before. Second, the usual CS-term of the form $F \wedge A - (1/3)A \wedge A \wedge A$ does *not* exist in our third-rank field strength (1.2b). This feature is different from the known tensor multiplet in the *Abelian* case [25][26]. Third, there are different non-conventional CS-terms in the field strengths $G_{abc}{}^I$ and $H_{ab}{}^I$. For these reasons, even the usual basic relationship for the scalar superfield $L$:

$$\left[ \nabla_\alpha, \overline{\nabla}_{\dot\beta} \right] L = c_1 \left( \sigma^{cde} \right)_{\alpha\dot\beta} G_{cde} + c_2 \, \mathrm{tr}\,(W_\alpha \overline{W}_{\dot\beta}) \tag{4.1}$$

does *not* hold. This is because the $G$-term on the right side is supposed to carry the adjoint index, while the second $W\overline{W}$-term does *not*, due to the trace taken.

One might think that the already-established 'off-shell' prepotential formulation [17][25][26] should be applicable to *any interactions*. However, such an expectation is *not* valid, because we are dealing with a tensor multiplet with an *adjoint index*, which is beyond the scope of the conventional prepotential formulation for a tensor multiplet as a *singlet* of any gauge group. This is the reason why even off-shell prepotential formulation for the *Abelian* tensor multiplet does *not* work in the *non-Abelian* case.

At the present time, we do not know how to overcome obstructions against an *off-shell* prepotential superfield formulation. The only way we can proceed is to rely on superspace Bianchi identities, as we have performed in the previous section, that can guarantee the consistency of our component formulation in section 2.

## 5.  Generating  $\mathcal{N} = 2$  and  $\mathcal{N} = 3$  Flows of Generalized KdV Eqs.

As applications of our SDNATM, we give the examples of embedding  $\mathcal{N} = 2$[7]  and  $\mathcal{N} = 3$  flows of generalized KdV eqs.  To this end, we perform the DR from the original  $D = 2+2$  into  $D = 1+1$.  For the original  $D = 2+2$, we use the coordinates  $(z, x, y, t)$  with the metric for  $D = 2 + 2$  [8]:

$$ds^2 = +2dzdx + 2dydt \quad . \tag{5.1}$$

The final  $D = 1 + 1$  has the coordinates  $(x, t)$.  For simplicity sake, we truncate all the fermionic fields:  $\lambda^I \overset{*}{=} \chi^I \overset{*}{=} \rho^I \overset{*}{=} 0$.  We now see that the SD condition (1.5d) on  $F$  is

$$F_{xt} \overset{*}{=} 0 \quad , \tag{5.2a}$$

$$F_{yz} \overset{*}{=} 0 \quad , \tag{5.2b}$$

$$F_{zx} \overset{*}{=} F_{ty} \quad , \tag{5.2c}$$

with  $\epsilon^{zxyt} = +1$.  Following the prescription in [8][24], we regard the YM filed components  $A_x$  and  $A_t$  in  $D = 2$  as *pure gauge*:

$$A_x \overset{*}{=} A_t \overset{*}{=} 0 \quad , \tag{5.3}$$

due to (5.2a).  We also require the independence of all the quantities on the  $y$  and  $z$-coordinates:  $\partial_y \overset{*}{=} 0, \ \partial_z \overset{*}{=} 0$, so (5.2b) and (5.2c) are equivalent to

$$[P, B] \overset{*}{=} O \quad , \tag{5.4a}$$

$$\dot{P} + B' \overset{*}{=} O \quad , \tag{5.4b}$$

where  $P \equiv A_y, \ B \equiv A_z$, and their *prime* and *dot* denote respectively the derivatives  $\partial_x \equiv \partial/\partial x$  and  $\partial_t \equiv \partial/\partial t$.

There are two remarks for the SD condition (1.5c):  First, since this SD shares the same index structure with (1.5d), we have the conditions parallel to (5.2):

$$H_{xt} \overset{*}{=} 0 \quad , \quad H_{yz} \overset{*}{=} 0 \quad , \quad H_{zx} \overset{*}{=} H_{ty} \quad . \tag{5.5}$$

---

[7] We use the symbol  $\mathcal{N}$  for these flows, in order to distinguish them from the number  $N$  of supersymmetries.

Second, the $gB$-term in the field strength $H$ (1.1c) can absorb the first gradient terms $\nabla C$, so that the $C$-field has no longer a dynamical field as a Stueckelberg field. So (5.5) is equivalent to

$$H_{xt} \overset{*}{=} B_{xt} \overset{*}{=} 0 \quad , \qquad H_{yz} \overset{*}{=} B_{yz} \overset{*}{=} 0 \quad , \qquad H_{zx} \overset{*}{=} B_{zx} \overset{*}{=} H_{ty} \overset{*}{=} B_{ty} \quad , \qquad (5.6)$$

where we put $g = 1$ from now on for simplicity.

We now perform the DR of the duality (1.5b). Using also equations above, we get

$$\varphi' \overset{*}{=} + B'_{yt} + \dot{B}_{xy} \quad , \qquad (5.7a)$$

$$\dot{\varphi} \overset{*}{=} + B'_{tz} + \dot{B}_{zx} \quad , \qquad (5.7b)$$

$$[P, \varphi] \overset{*}{=} - [B, B_{xy}] - [P, B_{zx}] \quad , \qquad (5.7c)$$

$$[B, \varphi] \overset{*}{=} - [P, B_{tz}] - [B, B_{yt}] \quad . \qquad (5.7d)$$

For simplicity sake, we impose additional conditions

$$\varphi - B_{zx} \overset{*}{=} \varphi - B_{ty} \overset{*}{=} 0 \quad , \qquad B_{tz} \overset{*}{=} 0 \quad , \qquad (5.8)$$

so that (5.7b) and (5.7c) are satisfied. Eventually, (5.7) is simplified to

$$[P, X] + [B, Y] \overset{*}{=} O \quad , \qquad (5.9a)$$

$$X' - \dot{Y} \overset{*}{=} O \quad , \qquad (5.9b)$$

where $X \equiv +\varphi - B_{xz} \overset{*}{=} \varphi - B_{yt}$, $Y \equiv +B_{xy}$. After all, the duality conditions in (1.5) are reduced to the four equations in (5.4) and (5.9).

We next give some examples of integrable systems that are generalized by our SDNATM system. As the first example, we show that the $\mathcal{N} = 2$, $r = 3$ flow of the generalized KdV equations [21], i.e., the original KdV equation:[8]

$$4\dot{u} \overset{\cdot}{=} + u''' + 6uu' = (u'' + 3u^2)' \qquad (5.10)$$

---

[8] We use the symbol $\overset{*}{=}$ for a field equation, or an equality valid upon field equation(s).

is embedded into (5.4) and (5.9). Our ansätze for $P$, $B$, $X$ and $Y$ are

$$P \overset{*}{=} \begin{pmatrix} 0 & 0 \\ u & 0 \end{pmatrix} \quad , \qquad B \overset{*}{=} -\frac{1}{4} \begin{pmatrix} 0 & 0 \\ u'' + 3u^2 & 0 \end{pmatrix} \quad , \tag{5.11a}$$

$$X \overset{*}{=} +\frac{1}{4} \begin{pmatrix} u'' + 3u^2 & 0 \\ 0 & 0 \end{pmatrix} \quad , \qquad Y \overset{*}{=} \begin{pmatrix} u & 0 \\ 0 & 0 \end{pmatrix} \quad . \tag{5.11b}$$

Now eq. (5.4) is easily satisfied by these $P$ and $B$. As for (5.9b), it generates (5.10). As for (5.9a), the only non-trivial component is its 21-component, which also vanishes as $(1/4)u(u'' + 3u^2) - (1/4)u(u'' + 3u^2) = 0$. Note that this is a non-trivial result, because each of the commutators $[P, X]$ and $[B, Y]$ is non-zero. This is also the reflection of the non-Abelian commutators in our SDNATM system, in particular, the non-Abelian couplings of TM to YM-field.

Compared with the SDSYM case [24], where $P$ and $B$ were just $1 \times 1$ matrices, our present system is less trivial, because of the new SD conditions (5.9). In our present SDNATM system, the $B$ and $P$-matrices are less trivial $2 \times 2$-matrices, but still the embedding is rather simple. Also, our embedding is relatively simpler, compared with [22][23], in which a sophisticated $H$ or $Q$-matrix is needed. Our present SDNATM system is simpler but still non-trivial at the same time. This seems to be the result of the simplification played by the new TM, showing the advantage of SDNATM system.

We next repeat a similar prescription for $\mathcal{N} = 3$, $r = 4$ flow of generalized KdV equations [21]:

$$3\dot{u}_2 \overset{.}{=} -u_2^{[4]} + 2u_3''' - (u_2^2)'' + 4(u_2 u_3)' \quad , \tag{5.12a}$$

$$9\dot{u}_3 \overset{.}{=} -2u_2^{[5]} + 3u_3^{[4]} - 6u_2 u_2''' - 12u_2' u_2'' - 4u_2^2 u_2' + 6(u_2 u_3')' + 6(u_3^2)' \quad , \tag{5.12b}$$

where $[n]$ stands for the $n$-th derivative by $\partial_x$. These are re-expressed as

$$3\dot{u}_2 \overset{.}{=} \left[ -u_2''' + 2u_3'' - (u_2)^2 + 4u_2 u_3 \right]' \equiv 3\left[ f(u_2, u_3) \right]' \quad , \tag{5.13a}$$

$$9\dot{u}_3 \overset{.}{=} \left[ -2u_2^{[4]} + 3u''' - 3(u_2')^2 - 6u_2 u_2'' - \frac{4}{3}u_2^3 + 6u_2 u_3' + 6(u_3)^2 \right]' \equiv +9\left[ g(u_2, u_3) \right]' \quad . \tag{5.13b}$$

Our ansätze for $P$, $B$, $X$ and $Y$ are in terms of $4 \times 4$ matrices are

$$P \overset{*}{=} \begin{pmatrix} O & O \\ \mathcal{U} & O \end{pmatrix} \quad , \qquad B \overset{*}{=} \begin{pmatrix} O & O \\ -\mathcal{F} & O \end{pmatrix} \quad , \qquad X \overset{*}{=} \begin{pmatrix} \mathcal{F} & O \\ O & O \end{pmatrix} \quad , \qquad Y \overset{*}{=} \begin{pmatrix} \mathcal{U} & O \\ O & O \end{pmatrix} \quad , \tag{5.14a}$$

$$\mathcal{F} \equiv \begin{pmatrix} f & 0 \\ g & f \end{pmatrix} \quad , \quad \mathcal{U} \equiv \begin{pmatrix} u_2 & 0 \\ u_3 & u_2 \end{pmatrix} \quad , \tag{5.14b}$$

where $f$ and $g$ are given in (5.13), while $\mathcal{F}$ and $\mathcal{U}$ are $2 \times 2$ matrices. We can easily show that all the conditions (5.4) and (5.9) are satisfied by these ansätze. In particular, the key relationship is the commutativity $[\mathcal{U}, \mathcal{F}] = O$. It seems that this kind of patterns can be repeated for higher hierarchies with larger $\mathcal{N}$ and $r$ for generalized KdV equations [21].

We have seen that the lower flows of generalized KdV equations [21] can be embedded into our SDNATM system. The most important ingredient is that the non-Abelian feature of TM is involved into the non-trivial embedding of these KdV equations, *via* commutators such as $[P, X]$ or $[B, Y]$. Even though the presence of the TM seems to complicate the system, it simplifies the matrices of $B$ and $P$ compared with SDYM system [22][23], where a more complicated $H$-matrix is needed. The embedding of KdV equations reveals the advantage of our SDNATM system over SDSYM system [11][12].

## 6. Relationship with N = 2 SDSYM

We can see that our system of SDNATM produces the same set of bosonic field equations as those by $N = 2$ SDSYM theory in $D = 2 + 2$ with the field content $(A_\mu{}^I, \lambda_i{}^I, T^I)$, where $i = 1, 2$ is the index for $N = 2$ supersymmetry. Each of $\lambda_1{}^I$ and $\lambda_2{}^I$ are Majorana-Weyl spinor with negative chirality, and $T^I$ is a real scalar in the adjoint representation [12]. As shown in [12], when the DR into $D = 1 + 1$ is performed, the set of bosonic conditions from $N = 2$ SDSYM are (5.4) and

$$[B, T'] + [P, \dot{T}] \stackrel{*}{=} 0 \quad , \tag{6.1}$$

We can show that the condition in (6.1) is equivalent to (5.9) arising in our SDNATM. Let $U(x, t)$ be a scalar defined by

$$U(x, t) \equiv + \int_{t_0}^t d\tau \, X(x, \tau) \quad , \tag{6.2}$$

so that

$$X = \frac{\partial U}{\partial t} = \dot{U} \quad , \quad X' = \frac{\partial^2 U}{\partial x \, \partial t} \quad . \tag{6.3}$$

Integrating (5.9b) over time, we get

$$Y = + \int_{t_0}^{t} d\tau \, X'(x, \tau) = + \int_{t_0}^{t} d\tau \, \frac{\partial^2 U(x, \tau)}{\partial x \, \partial \tau}$$

$$= + \int_{t_0}^{t} d\tau \, \frac{\partial}{\partial \tau} \left[ \frac{\partial(x, \tau)}{\partial x} \right] = + \frac{\partial U(x, t)}{\partial x} = + U' \quad \Longrightarrow \quad Y = U' \quad . \tag{6.4}$$

Then (5.9a) is expressed in terms of $U$ as

$$[B, U'] + [P, \dot{U}] \overset{*}{=} 0 \quad . \tag{6.5}$$

This is nothing but (6.1) with $T$ replaced by $U$. In other words, our $N = 1$ SDNATM generates the same bosonic conditions as $N = 2$ SDSYM theory [12], despite simple supersymmetry $N = 1$ in our system instead of extended $N = 2$ in [12].

This result is natural, because even though we have only $N = 1$ supersymmetry, since the system of SDNATM is larger than $N = 1$ SDSYM, the enlargement resulted in the equivalence to the *enhanced* supersymmetry from $N = 1$ to $N = 2$, when a DR into $D = 1 + 1$ is performed.

## 7. Concluding Remarks

In this paper, following the recent successful formulations of non-Abelian tensors [1][2][4][7], we have first presented the component formulation of an $N = 1$ SDNATM theory with non-trivial couplings to YMVM. Our system has three multiplets (i) YMVM $(A_\mu{}^I, \lambda^I)$, (ii) NATM $(B_{\mu\nu}{}^I, \chi^I, \varphi^I)$, and ECVM $(C_\mu{}^I, \rho^I)$. Similarly to our recent formulation of $N = 1$ TM in $D = 3 + 1$ [7], we need the three multiplets of TM, VM and ECVM. In particular, the ECVM is indispensable for the consistent couplings of TM to VM. The usual YM field strength $F_{\mu\nu}{}^I$, and the field strength $H_{\mu\nu}{}^I$ of the extra compensator vector $C_\mu{}^I$ should be also self-dual, in order to accomplish the total consistency.

As independent confirmation, we have also given superspace re-formulation, showing the consistency with the component formulation. Our superfield formulation is *on-shell* formulation based on the fundamental superfields VM $(A_a{}^I, \overline{\lambda}_{\dot{\alpha}}{}^I)$, SDTM $(B_{ab}{}^I, \chi_\alpha{}^I, \varphi^I)$ and ECVM $(C_a{}^I, \overline{\rho}_{\dot{\alpha}}{}^I)$. Even though this is *on-shell* formulation without prepotentials, this is the very first formulation in superspace for a self-dual tensor multiplet. This situation is similar to our superspace formulation in [7] as the very first superspace formulation for a non-Abelian tensor multiplet. As for the *off-shell* formulation, we leave it to future studies, due to non-trivial field strengths involved, and the prepotential formulation would be very involved.

To our knowledge, combining non-Abelian TM with SD, $N = 1$ supersymmetry and integrable models has not been entertained in the past literature. We have given not only the component formulation, but also superspace re-formulation for the first time, as supporting evidence for the total consistency. The successful coupling of a tensor field with the adjoint index of a non-Abelian gauge group is based on the extra terms in the field strengths $G$ and $H$ inspired from the recent works [1][2][4][7]. In particular, the extra compensator vector $C_\mu{}^I$ in the ECVM serves as the Stueckelberg field to be absorbed into the longitudinal component of $B_{\mu\nu}{}^I$. This seems to imply that the Stueckelberg mechanism is inevitable for avoiding inconsistency by the naïve couplings of TM. This feature is common both to $D = 3 + 1$ [7] and $D = 2 + 2$ space-time dimensions.

As applications, we have also given the examples of generalized KdV equations for the $\mathcal{N} = 2$, $r = 3$ and $\mathcal{N} = 3$, $r = 4$ flows. Our new duality symmetry (1.5a) and (1.5b) for the TM provides a set of non-trivial conditions (5.9), in addition to those with SDSYM with a pure VM [12]. The embeddings into the $P$ and $B$-matrices given in [24] were rather trivial, because they were only $1 \times 1$-matrices, while in our present case, the matrices $P$ and $B$ are at least $2 \times 2$-matrices.

Our SDNATM system has much simpler embedding configurations, compared with SDSYM theories [12]. For example, we have seen that our original SDNATM has only $N = 1$ supersymmetry, it generates in $D = 1 + 1$ the same set of conditions produced by $N = 2$ SDSYM [12]. Of course, the price to be paid is the introduction of the new set of duality symmetry (1.5a) and (1.5b) resulting in (5.9), but it is compensated by the simplification of embedding. Our configurations are much simpler and more straightforward than [22][23], but still non-trivial for generalized KdV equations [21].

The work presented here initiates new directions of research on supersymmetric duality symmetry in $D = 2 + 2$, as well as in higher dimensions. To be specific, we can potentially generalize our result beyond 4D for *generalized* SD [27][28][29]. as follows. Our SD (1.5) is generalized to higher-dimensions in $D$ as

$$F_{\mu\nu}{}^I \overset{*}{=} + \tfrac{1}{2}\phi_{\mu\nu}{}^{\rho\sigma}F_{\rho\sigma}{}^I \ , \tag{7.1a}$$

$$G_{\mu\nu\rho}{}^I \overset{*}{=} - \phi_{\mu\nu\rho}{}^{\sigma}\nabla_\sigma\varphi^I \ , \tag{7.1b}$$

$$H_{\mu\nu}{}^I \overset{*}{=} + \tfrac{1}{2}\phi_{\mu\nu}{}^{\rho\sigma}H_{\rho\sigma}{}^I \ , \tag{7.1c}$$

with an appropriate constant $\phi_{\mu\nu}{}^{\rho\sigma}$, *e.g.*, the octonion structure constant [30] in 8D for a reduced holonomy $SO(7) \subset SO(8)$ [27][28][29]. This kind of generalizations especially with non-Abelian tensors has become within our reach, after the successful formulations of non-Abelian tensors in 4D [1][2][4][7].

Technical details aside, the conceptual lessen we can learn from our present result is as follows. The original Atiyah-Ward conjecture [9] was that all the lower-dimensional bosonic integrable systems in $D \le 3$ are generated by SDYM theory in $D = 2 + 2$. In 1990's, this conjecture was further supersymmetrized to SDSYM systems [11][14][12]. Now it is the next natural step to consider the generalization of a SDYM to a SD non-Abelian tensor. We can further consider the higher-dimensional generalization of SD in 4D to 7D or 8D, based on the so-called *reduced holonomy* [27][28][29], as in (7.1). In other words, theories evolve from Abelian groups to non-Abelian groups, from non-supersymmetric to supersymmetric systems, from vectors to tensors, and from $D = 4$ to $D \ge 5$. It is clear that our present result has historical implication contributing to the past accomplishments [27][28][29], as well as inducing future applications. We also emphasize that the generalization to non-Abelian tensor has been made possible, only after the success of NATM in $D = 3 + 1$ [1][2][4][7].

We are indebted to the referees of this paper for important suggestions to improve the paper. This work is supported in part by Department of Energy grant # DE-FG02-10ER41693.